\documentclass[
reprint,
superscriptaddress,
tightenlines,
aps,
prl,
floatfix
]{revtex4-1}

\usepackage{amsmath}
\usepackage{amssymb}
\usepackage{amsthm}
\usepackage{chngcntr}
\usepackage{dcolumn}
\usepackage{float}
\usepackage{graphicx}
\usepackage[colorlinks=true, allcolors=blue]{hyperref}
\usepackage{lipsum}
\usepackage{tabularray}
\begin{document}

\title{Higher-order Ising model on hypergraphs}

\author{Thomas Robiglio}
\affiliation{Inverse Complexity Lab, IT:U Interdisciplinary Transformation University Austria, 4040 Linz, Austria}
\affiliation{Department of Network and Data Science, Central European University, Vienna, Austria}
\author{Leonardo Di Gaetano}
\affiliation{Department of Network and Data Science, Central European University, Vienna, Austria}
\author{Ada Altieri}
\affiliation{Laboratoire Matière et Systèmes Complexes (MSC), Université Paris Cité, CNRS, 75013 Paris, France}
\author{Giovanni Petri}
\affiliation{NPLab, Network Science Institute, Northeastern University London, London, UK}
\affiliation{CENTAI Institute, Turin, Italy}
\author{Federico Battiston}
\email{battistonf@ceu.edu}
\affiliation{Department of Network and Data Science, Central European University, Vienna, Austria}
\date{\today} 

\begin{abstract} 
Non-dyadic higher-order interactions affect collective behavior in various networked dynamical systems. Here we discuss the properties of a novel Ising model with higher-order interactions and characterize its phase transitions between the ordered and the disordered phase. By a mean-field treatment, we show that the transition is continuous when only three-body interactions are considered but becomes abrupt when interactions of higher orders are introduced. Using a Georges-Yedidia expansion to go beyond a naïve mean-field approximation, we reveal a quantitative shift in the critical point of the phase transition, which does not affect the universality class of the model.
Finally, we compare our results with traditional $p$-spin models with many-body interactions. Our work unveils new collective phenomena on complex interacting systems, revealing the importance of investigating higher-order systems beyond three-body interactions.
\end{abstract}
\maketitle

Many complex systems are characterized by non-pairwise interactions among the system's units~\cite{battiston2020networks}. 
 Taking into account the higher-order structure of networks has led to the discovery of new phenomena and collective behaviors across a wide range of dynamical processes~\cite{battiston2021physics}, including contagion~\cite{iacopini2019simplicial,kim2024higher,burgio2024triadic,ferraz2023multistability}, diffusion~\cite{schaub2020random,carletti2020random,di2024dynamical}, synchronization~\cite{millan2020explosive,lucas2020multiorder,gambuzza2021stability, zhang2023higher,zhang2024deeper}, evolutionary dynamics in social dilemmas~\cite{alvarez2021evolutionary, civilini2024explosive} and ecology~\cite{grilli2017higher, bairey2016high}.

The Ising model is one of the simplest models to display a phase transition, universality, and complex phenomenology~\cite{brush1967history}. Originally introduced to study the order-disorder transition in lattices for ferromagnets, it has been extended to consider general interaction structures modeled as complex networks~\cite{leone2002ferromagnetic, dorogovtsev2002ising, bianconi2002mean}, finding a rich phenomenology where the presence and nature of the phase transition depend on the specific shape of the degree distribution.
Many-body interactions have been considered before in spin models, under the name of $p$-spin models~\cite{derrida1980random}.
Yet, the straightforward extension of the Ising model to higher-order interactions provided by the $p$-spin model breaks the $\mathbb{Z}_1$ of the pairwise model when odd $p$ interactions are considered and assigns the same energy to different spin configurations when there are even $p$ interactions. 
Here we study the phase transitions of a novel spin model with dyadic and group interactions which we originally proposed in Ref.~\cite{robiglio2024synergistic} to overcome the limitations mentioned above of the traditional $p$-spin approaches.
\begin{figure}
    \centering
    \includegraphics{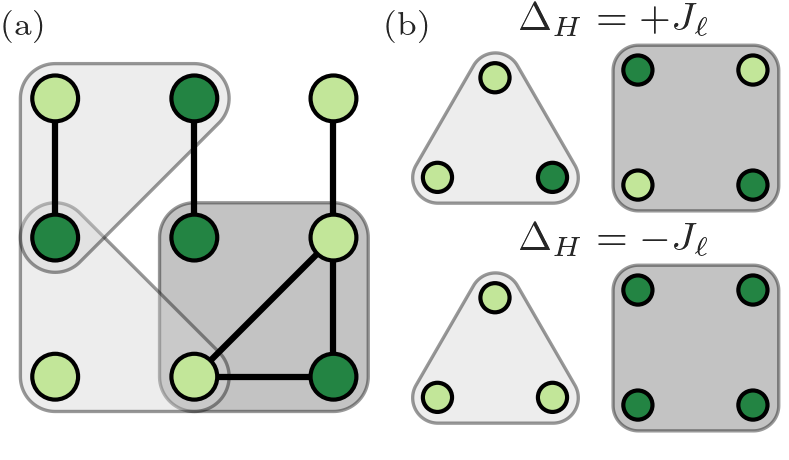}
    \caption{Illustration of the higher-order Ising model. (a) The model is defined on a hypergraph, the nodes interact through edges and hyperedges of different orders. (b) Hyperedges of order $\ell$ contribute to the energy with a negative term $\Delta_{\ell}=-J_{\ell}$ if all the nodes contained in them are aligned, otherwise the contribution has the opposite sign.}
    \label{fig:fig1}
    \vspace{-1.em}
\end{figure}

We develop a mean-field treatment of this model and discuss its emergent behaviors in the case of arbitrarily structured and heterogeneous higher-order interactions modeled by hypergraphs. 
Our model displays complex behavior, with the nature of the disorder-order phase transition depending on the maximum order of the many-body interactions: continuous for three-body interactions, and discontinuous for higher orders. 
We perform a high-temperature expansion of the free energy of our model to go beyond the mean-field approximation, showing that sparse connectivity induces a correction in the critical point marking the onset of the phase transition, without affecting the universality class of the model.
We also compare our results with traditional ferromagnetic $p$-spin models in the literature.

\paragraph{Model.}
We consider the Ising model on higher-order networks we introduced in Ref.~\cite{robiglio2024synergistic}.
This model is defined on a hypergraph $\mathcal{H}=(V,E)$ with $|V|=N$ nodes and hyperedges $\{\sigma\}=E$---subsets of elements of $V$---of order (number of nodes participating in the hyperedge minus 1) up to $\ell_{\rm max}$.
To each node $i \in V$ we associate a binary state variable $s_i\in \{\pm 1\}$, corresponding to the spin state of the node, which can either be up ($s_i=+1$) or down ($s_i=-1$).
The global state of the system is controlled by the Hamiltonian:
\begin{equation}
\label{eq:all_orders_conserved_symmetry}
    H^{\rm CS} = - h \sum_{i} s_i
     - \sum_{\ell=1}^{\ell_{\text{max}}} J_{\ell} 
    \sum_{\{\sigma \in \mathcal{H}:|\sigma|=\ell\}}
    \left[2 \bigotimes_{i\in \sigma}s_i -1 \right]
\end{equation}
where:
\begin{equation}
    \bigotimes_{i=1}^n s_i =\delta\left(s_1,...,s_n\right)
    =\begin{cases}
    1 & \text{if\;} s_1=...=s_n \\
    0 & \text{otherwise}
    \end{cases}
\end{equation}
is the Kronecker delta for an arbitrary number of binary arguments.
This higher-order generalization of the dyadic model echoes previous works extending pairwise contagion models to higher-order networks, where susceptible nodes can get infected via additional group mechanisms if all the other nodes participating in a given hyperedge are infected~\cite{iacopini2019simplicial}.
Note that if the spin variables take discrete values $s_i\in\{0,1,...,q-1\}$ Eq.~\eqref{eq:all_orders_conserved_symmetry} can be modified to describe a generalization to arbitrary orders of interaction of the standard Pott's model~\cite{wu1982potts}.
An alternative extension of the Ising model to account for many-body interactions is described by the Hamiltonian:
\begin{equation}
\label{eq:all_orders_broken_symmetry}
H^{\rm BS} =-h\sum_{i} s_i - \sum_{\ell=1}^{\ell_{\rm max}}J_{\ell} \sum_{\{\sigma \in \mathcal{H}:|\sigma|=\ell\}} \prod_{i\in \sigma} s_i
\end{equation}
This model is often referred in the literature as $p$-spin model (with $p=\ell+1$) and has been extensively studied in fully connected~\cite{barra2009notes}, diluted~\cite{newmanglassy1999, franz2001, agliari2011notes}, and disordered systems~\cite{derrida1980random}.
As already noted in the first analysis of the Ising model with three-body interactions, when there are interactions between an odd number of spins, the extension provided by Eq.~\eqref{eq:all_orders_broken_symmetry} breaks the parity symmetry (for this reason we refer to this model as \textit{broken symmetry} model, BS, and to our model described by Eq.~\eqref{eq:all_orders_conserved_symmetry} as \textit{conserved symmetry}, CS) under spin flip at all sites of the dyadic model without magnetic field~\cite{merlini1973symmetry}.
This can be easily understood considering the energy of a system constituted by three spins connected in a 2-hyperedge. 
There are two states with all spins aligned: all spins pointing up and all spins pointing down.
These two configurations are symmetric upon flipping all spins, yet their energies when computed using Eq.~\eqref{eq:all_orders_broken_symmetry} are different, favoring the state with spins pointing up. 
Moreover, when considering orders $\ell>2$ Eq.~\eqref{eq:all_orders_broken_symmetry} assigns the same energy to very different spin configurations when quadratic terms are present. 
For example, considering a plaquette interaction of four spins this model will assign the same (ground-state) energy to the configurations in which all the spins are aligned and to the configurations in which there are pairs of spins pointing in the same direction.
The CS model was originally formulated to overcome these inherent limitations of the traditional ferromagnetic $p$-spin model.

\paragraph{Homogeneous mean-field}
We develop a homogeneous mean-field treatment of the model based on two assumptions.
Firstly, we will write the spin state at site $i$ as:
\begin{equation}
\label{eq:mf_trick_1}
    s_i = \langle s_i \rangle + \Delta s_i = \langle s_i \rangle + (s_i - \langle s_i \rangle)
\end{equation}
We simplify the coupling terms appearing in the Hamiltonians by neglecting all terms that are second-order in the fluctuations.
Secondly, we assume that the expectation value of the spin-state is uniform in the entire system:
\begin{equation}
\label{eq:mf_trick_2}
\langle s_i \rangle = m \;\; \forall i
\end{equation}
where the magnetization $m=\sum_i s_i/N$ is the order parameter of the system, taking values in the interval $[-1,+1]$.
The pairwise Ising model (corresponding to our model and to the $p$-spin model with $\ell_{\rm max}=1$) with no magnetic field, when varying the inverse temperature $\beta$, undergoes a second-order phase transition between a disordered ($m=0$ for $\beta=0$) and an ordered phase ($|m|=1$ for $\beta \to \infty$).
This phase transition can alternatively be expressed by fixing the inverse temperature $\beta$ and varying the coupling strength $J_1$.
Using Eqs.~\eqref{eq:mf_trick_1}-\eqref{eq:mf_trick_2}, we can write the product of two spins appearing in the coupling term of the pairwise Ising model as $m(s_i+s_j)-m^2$.
In Eq.~\eqref{eq:all_orders_conserved_symmetry} we will have the same term for $\ell=1$ as for $s_i,s_j \in \{-1,+1\}$ we have the identity:
\begin{equation}
\label{eq:identity_delta_product}
    2\delta(s_i,s_j)-1\equiv s_is_j \Longleftrightarrow \delta_{s_i,s_j} \equiv \frac{s_is_j+1}{2}
\end{equation}
Considering interactions of arbitrary order $\ell$ we make use of the identity:
\begin{equation}
    \bigotimes_{i=1}^n s_i\equiv \prod_{j=2}^n\delta(s_1, s_j)
\end{equation}
We can thus write the many-body terms appearing in the Hamiltonian Eq.~\eqref{eq:all_orders_conserved_symmetry} as:
\begin{equation}
\label{eq:general-conserved-sim}
2\bigotimes_{i=1}^n s_i =  \frac{1}{2^{n-1}}\left[\sum_{\alpha=2}^{2 \left\lfloor \frac{n}{2} \right\rfloor} \prod_{i=1}^{\alpha} s_{i}+1\right]
\end{equation}
where $2 \left\lfloor \frac{n}{2} \right\rfloor$ is the largest even number smaller or equal to $n$.
For instance, if $n$ = 5, $2 \left\lfloor \frac{n}{2} \right\rfloor = 4$ and the latter equation will contain products of spins up to size $4$.
Considering the product of $\alpha$ spins appearing in Eq.~\eqref{eq:general-conserved-sim}, in the mean-field approximation we have:
\begin{equation}
\label{eq:prod-spin}
    \prod_{i =1}^{\alpha} \ s_i  \simeq m^{\alpha-1} \sum _{i=1}^{\alpha} s_i  - \left(\left(\alpha-1\right) \ m^{\alpha} \right)
\end{equation}
Inserting Eq.~\eqref{eq:prod-spin} into Eq.~\eqref{eq:general-conserved-sim} allows us to write explicitly the decoupling of the Kronecker delta in terms of the single nodes states and powers of the magnetization.
At all orders, under the mean-field assumption we decouple all spins obtaining a fully decoupled Hamiltonian $H(m)=-h_{\rm eff}\sum_i s_i$.
As the coupling parameters $J_{\ell}$ appear in $H(m)$ multiplied with the corresponding generalized average degree $\langle d_{\ell} \rangle$, we introduce the rescaled coupling parameters $\gamma_{\ell}=J_{\ell}\langle d_{\ell} \rangle$.
\paragraph{Constrained free energy.}
We write the partition function---the normalization constant in the Boltzmann distribution over spin configurations $\{\mathbf{s}\}$---as:
\begin{equation}
\label{eq:mf_partition_function}
    Z=\sum_{\{\mathbf{s}\}}\exp[-\beta H(\mathbf{s})] \simeq \sum_m g(m)\exp[-\beta H(m)]
\end{equation}
where $g(m)$ counts the number of configurations with magnetization $m$.
Using Stirling's approximation we can easily write $\log g(m)$ in the familiar form of a binary entropy.
We can then introduce the constrained free energy density:
\begin{equation}
\label{eq:constrained_free_energy_density}
    f(m)=\frac{H(m)}{N}-\frac{\log g(m)}{\beta N}
\end{equation}
The minimization of the constrained free energy reduces to the implicit equation:
\begin{equation}
\label{eq:ising_paramagnet}
    m=\tanh \left[\left(\beta h_{\rm eff}(m)\right)\right]
\end{equation}

\begin{figure}
    \centering
    \includegraphics{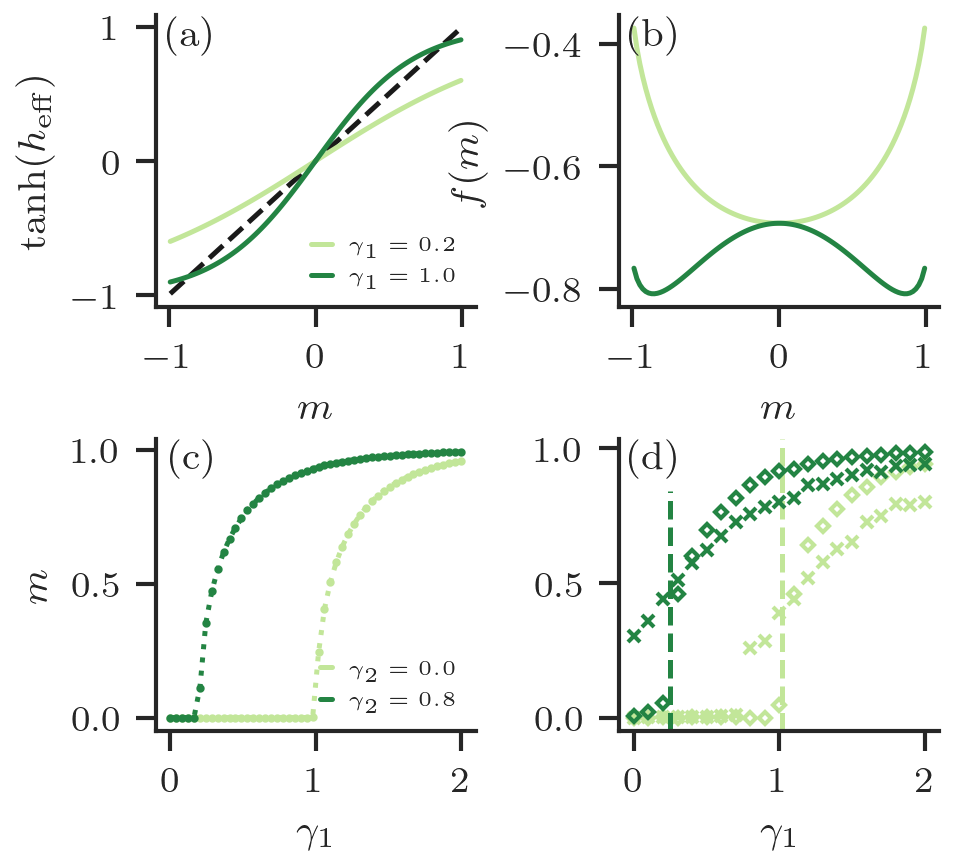}
    \caption{(a) Shape of the right-hand side of Eq.~\eqref{eq:ising_paramagnet} and (b) functional form of the mean-field constrained free energy density in the case in which $h=0$, $\ell_{\rm max}=2$, and $\gamma_2$. The dashed line in (a) is the bisector, corresponding to the left-hand side of the equation of state. (c) Phase transition in the $m(\gamma_1)$ phase space. (d) Numerical results on random hypergraphs with $N = 1000$ nodes and generalized degrees $\langle k_1\rangle =20$ $\langle k_2\rangle =6$ (squares) and uncorrelated scale-free hypergraphs with $N=1000$ and $\alpha_1=\alpha_2=3$ (crosses). The dashed lines correspond to the critical pairwise coupling found in the mean-field approximation. }
    \label{fig:fig2}
    \vspace{-1em}
\end{figure}

\paragraph{Mean-field results with three-body interactions.}
When 3-body interactions are considered, the novel (CS) model displays significant differences in the solutions of the equation of state Eq.~\eqref{eq:ising_paramagnet} and in the energy landscape with respect to the $p$-spin (BS) model (Fig.~\ref{fig:fig2}a,b).
We see that the functional form of the constrained free energy $f(m)$ is symmetric for the CS model.
As expected, in the CS model the ferromagnetic ground state is degenerate \textit{i.e.} we have two symmetric minima corresponding to $\pm m$. 
In contrast, in the BS model we have only one ferromagnetic solution with positive magnetization (see SM) due to the symmetry breaking induced by the coupling between an odd number of spins. 
We see in Fig.~\ref{fig:fig2}c that keeping fixed $\gamma_2$ and $\beta$, and varying $\gamma_1$ the system undergoes a continuous phase transition between the two phases mentioned above.
This is not observed in the $p$-spin model where the introduction of many-body interactions ($\gamma_2>0$) gives an abrupt phase transition~\cite{franz2001}.
We have validated these results through Monte Carlo simulations~\cite{newman1999monte} on random hypergraphs~\cite{landry2020effect, landry2023opinion} (Fig.~\ref{fig:fig2}d).
When the degree and generalized degree distributions are sharply peaked around their average values (Erdős–Rényi-like hypergraphs) we have very good agreement between the mean-field analytics and the numerical results.
When we consider heterogeneous structures---configurations hypergraphs in which power-laws with exponent $\alpha=3$ describe the degree and generalized degree distributions---the fat-tailed degree distributions produce a lowering of the magnetization threshold.
This lowering of the threshold for the onset of the phase transition echoes known results for pairwise dynamical models on scale-free networks, where the degree heterogeneity induces in finite systems small critical thresholds, which vanish as the system size is increased~\cite{pastor2002epidemic,bianconi2002mean}
\begin{figure}
    \centering
    \vspace{1em}
    \includegraphics{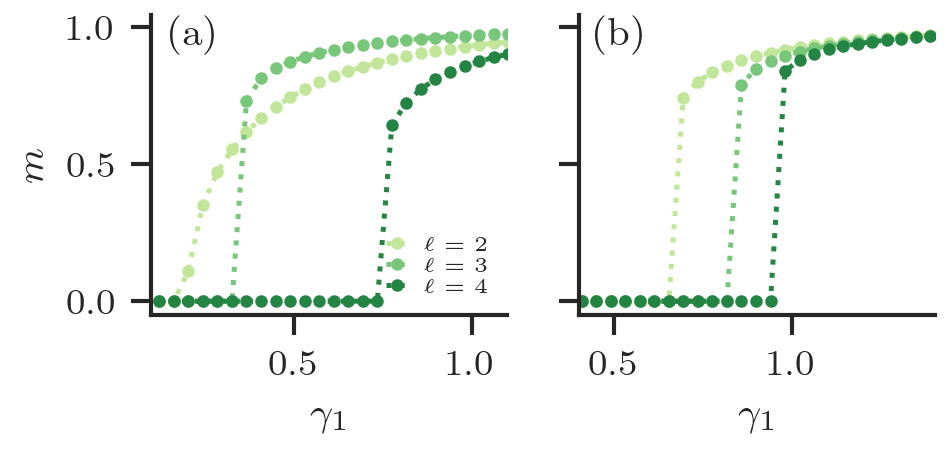}
    \caption{Phase transition in the $m(\gamma_1)$ phase space for the mean-field solution of the higher-order Ising model (a) and for the ferromagnetic $p$-spin model in the case with $h=0$, pairwise interactions controlled by $\gamma_1$ and group interactions of order $\ell\in \left[ 2,3,4\right]$ with fixed $\gamma_{\ell}=0.8$.}
    \label{fig:fig3}
    \vspace{-1em}
\end{figure}
\paragraph{Phase transition beyond three-body interactions.}
In the case $\ell_{\rm max}=2$ we have that the phase transition between the disordered and the magnetized phase is continuous, differently from what is observed in $p=3$-spin models where the transition is explosive.
This difference in the phase transition behavior between the two models vanishes once interactions of order $\ell \geq 3$ (4-body interactions and higher) are introduced (Fig.~\ref{fig:fig3}). 
The explosive transition results from the presence of powers of $m$ in the expression for $h_{\rm eff}$. 
In the BS model, interactions of order $\ell$ introduce a factor of $m^{\ell}$.
Similarly, in the CS model, interactions of order $\ell > 2$ introduce powers of $m$ in $h_{\rm eff}$. 
For example, considering a system with $\ell_{\rm max} = 3$ (4-body interactions), using Eqs.~\eqref{eq:general-conserved-sim}-\eqref{eq:prod-spin} we find the self-consistent equation for $m$ a term proportional to $m^2$, triggering an abrupt phase transition when the pairwise coupling term is varied.
\begin{figure}
    \centering
    \includegraphics{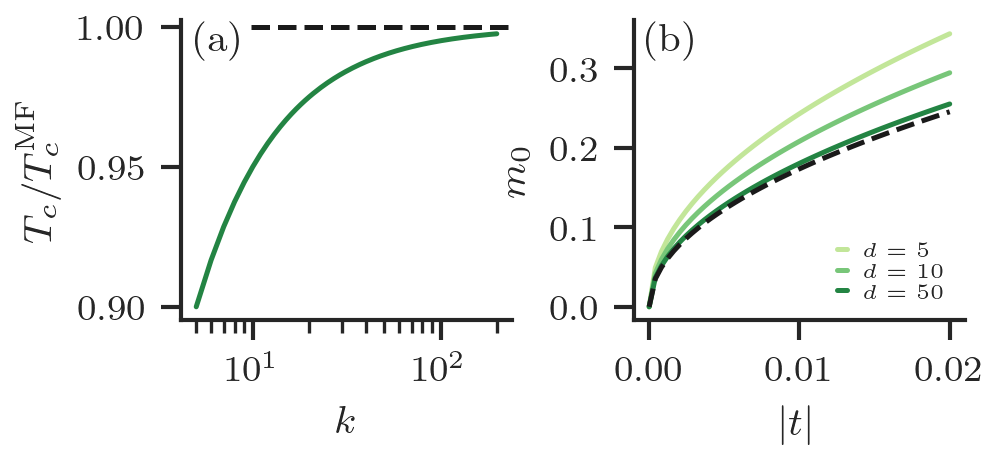}
    \caption{(a) Ratio between the critical temperature obtained with the G.-Y. expansion and the mean-field critical temperature for $d$-regular 2-hypergraphs. (b) Results for the spontaneous magnetization in the vicinity of the critical point for $d$-regular 2-hypergraphs with $J_2=1$. The dashed line is the mean-field result.}
    \label{fig:fig4}
    \vspace{-1em}
\end{figure}
\paragraph{Beyond the mean-field approximation.}
To improve on our mean-field estimation we can use high-temperature expansions of the free energy functional of our system at fixed order parameter~\cite{plefka1982convergence, georges1991expand, altieri2016}.
We perform a Georges-Yedidia (G.-Y.) expansion by defining for a general spin system with Hamiltonian $H$, a magnetization-dependent free energy functional:
\begin{equation}
    \mathcal{F}^{\beta}[\mathbf{m}]=\log \sum_{\{\mathbf{s}\}}\exp \left[-\beta H(\mathbf{s})+\sum_i \rho_i^{\beta}(S_i-m_i)\right]
\end{equation}
The Lagrange multipliers $\rho_i^{\beta}$ fix the magnetization at each site $i$ to their thermal expectation.
We then expand $\mathcal{F}^{\beta}[\mathbf{m}]$ around $\beta=0$ using a Taylor expansion:
\begin{widetext}
\begin{equation}
    \left. \mathcal{F}^{\beta}[\mathbf{m}]=\mathcal{F}^{\beta}[\mathbf{m}]\right|_{\beta=0}+\left.\frac{\partial \mathcal{F}^{\beta}[\mathbf{m}]}{\partial \beta}\right|_{\beta = 0}\beta + \frac{1}{2}\left.\frac{\partial^2 \mathcal{F}^{\beta}[\mathbf{m}]}{\partial \beta^2}\right|_{\beta = 0}\beta^2+...
\end{equation}
\end{widetext}
The derivatives of $\mathcal{F}^{\beta}[\mathbf{m}]$ evaluated at $\beta=0$ can be easily computed as spins are uncorrelated at infinite temperature: $\langle \prod_{i=1}^{\alpha} s_i\rangle|_{\beta=0}=\prod_{i=1}^{\alpha} m_i$.
Computing the first two terms gives the standard mean-field theory and higher-order derivatives provide corrections to the mean-field picture that are valid beyond the high-temperature regime. 
For example, for the Sherrington-Kirkpatrick model~\cite{sherrington1975solvable} the $\beta^2$ term gives the Onsager reaction term of the TAP equations~\cite{thouless1977solution}.
We perform this expansion for the CS model in the simple cases of fully connected systems and $d$-regular 2-hypergraphs (see SM for the detailed derivation).
In the fully connected case---as expected---the mean-field is exact in the thermodynamic limit: individual couplings vanish in the infinite system size and the expansion can be truncated after the second term.
In the case of $d$-regular hypergraphs with only three-body interactions, we can write the constrained free-energy density as:
\begin{widetext}
\begin{equation}
\label{eq:corrected_fm}
    f(m)=\log g(m) - \frac{\beta J_2d}{2}m^2-\frac{(\beta J_2)^2 d}{8}(1-m^2)^2+O(\beta^3)
\end{equation}
\end{widetext}
We explore the effect of the first correction term on the critical temperature $T_c$ and the spontaneous magnetization $m_0$---magnetization for $T\leq T_c$.
To find the critical temperature we expand Eq.~\eqref{eq:corrected_fm} in powers of $m$ and find the temperature for which the $m^2$ coefficient vanishes.
This gives:
\begin{equation}
    	T_c = J_2d\left(1-\frac{1}{2d}+O\left(\frac{1}{d^2}\right)\right)
\end{equation}
revealing the role of sparsity in the hypergraph structure (encoded by the higher-order degree $d$), which produces a lowering of the critical temperature (see Fig.~\ref{fig:fig4}a).
For large values of $d$, we recover the fully connected regime where the mean-field predictions are exact.
The spontaneous magnetization is obtained by expanding to order $m^3$ the minimization condition for $f(m)$.
As we see in Fig.~\ref{fig:fig4}b the $1/d$ expansion that we performed varies the amplitude of the critical scaling $m \sim \sqrt{-t}$---where $t$ is the reduced temperature $t=(T-T_c)/T_c$---but maintains the mean-field critical exponent.
\paragraph{Discussion.}
We have introduced a novel Ising-like model on higher-order networks, with both arbitrary structure and orders of interactions. 
By an approximate mean-field solution, we have shown that, when 3-body interactions are introduced, the model displays the emergence of a continuous transition toward an ordered phase. 
This phenomenology is different from traditional $p$-spin models, which are characterized by an abrupt transition. 
Our results are confirmed by numerical simulations on homogeneous structures, while heterogeneous hypergraphs anticipate the onset of the transition. 
When further orders of interactions are considered, our model recovers explosive behavior. 
Finally, we performed a high-temperature expansion of the model capturing corrections to the mean-field predictions of the behavior of the model by accounting for the effect of structure and the presence of feedback loops.
Our work unveils new collective phenomena in spin models on complex interacting systems, revealing the importance of studying higher-order interactions beyond three-body.

\bibliography{reference}

\end{document}


\title{Supplemental Material}

\renewcommand*{\citenumfont}[1]{S#1}
\renewcommand*{\bibnumfmt}[1]{[S#1]}

\setcounter{figure}{0}
\setcounter{table}{0}
\setcounter{equation}{0}
\setcounter{page}{1}
\setcounter{section}{0}

\makeatletter
\renewcommand{\thepage}{\roman{page}}
\renewcommand{\thefigure}{S\arabic{figure}}
\renewcommand{\theequation}{S\arabic{equation}}
\renewcommand{\thetable}{S\arabic{table}}

\setcounter{secnumdepth}{2}

\maketitle

\onecolumngrid

\vspace{-3em}

\section{Homogeneous mean-field treatment}
We present the full derivation of the homogeneous mean-field outlined in the main text. 
We carry out in parallel the derivation for the $p$-spin model (BS model) with Hamiltonian:
\begin{equation}
\label{eq:all_orders_broken_symmetry}
H=-h\sum_{i} s_i - \sum_{\ell=1}^{\ell_{\rm max}}J_{\ell} \sum_{\{\sigma \in \mathcal{H}:|\sigma|=\ell\}} \prod_{i\in \sigma} s_i
\end{equation}
and for the CS higher-order Ising model with Hamiltonian:
\begin{equation}
\label{eq:all_orders_conserved_symmetry}
    H = - h \sum_{i=1} s_i
     - \sum_{\ell=1}^{\ell_{\text{max}}} J_{\ell} 
    \sum_{\{\sigma \in \mathcal{H}:|\sigma|=\ell\}}
    \left[2 \bigotimes_{i\in \sigma}s_i -1 \right]
\end{equation}
where:
\begin{equation}
    \bigotimes_{i=1}^n s_i =\delta\left(s_1,...,s_n\right)
    =\begin{cases}
    1 & \text{if\;} s_1=...=s_n \\
    0 & \text{otherwise}
    \end{cases}
\end{equation}
is the Kronecker delta for an arbitrary number of binary arguments.
In the case in which $\ell_{\rm max}=2$ (\textit{i.e.} in the case in which we have dyadic and three-body interactions) we can write the two Hamiltonians as:
\begin{subequations}
\label{eq:three_body_hamiltonians}
\begin{align}
    H^{\rm BS} & = -h\sum_{i}s_i-J_1 \sum_{\left( i,j\right)}s_is_j -J_2\sum_{\left( i,j,k \right)} s_is_js_k
    \\
    H^{\rm CS} & = -h\sum_{i}s_i-J_1 \sum_{\left( i,j\right)} \left[2\delta(s_i,s_j)-1\right]
    -J_2 \sum_{\left( i,j,k\right)} \left[2\delta(s_i,s_j,s_k)-1\right]
\end{align}
\end{subequations}
where  $\left( i,j\right)$ and $\left( i,j,k\right)$ denote the two-body and the three-body connections in the hypergraph, respectively.

The homogeneous mean-field treatment of the two models is based on two assumptions.
Firstly, we will write the spin state at site $i$ as:
\begin{equation}
\label{eq:mf_trick_1}
    s_i = \langle s_i \rangle + \Delta s_i = \langle s_i \rangle + (s_i - \langle s_i \rangle)
\end{equation}
and we will then neglect all terms appearing in the Hamiltonians that are second-order in the fluctuations.
Secondly, we assume that the expectation value of the spin-state is uniform in the entire system:
\begin{equation}
\label{eq:mf_trick_2}
\langle s_i \rangle = m \;\; \forall i
\end{equation}

\subsection{Pairwise terms}
Using Eqs.~\eqref{eq:mf_trick_1}-\eqref{eq:mf_trick_2} we can write the product of two spins appearing in the pairwise term of $H^{\rm BS}$ as:
\begin{equation}
    \label{eq:two_body_term}
    s_is_j = \left[\langle s_i \rangle + (s_i - \langle s_i\rangle)\right]\left[\langle s_j \rangle + (s_j - \langle s_j \rangle)\right]\nonumber \\
    \simeq m(s_i+s_j)-m^2
\end{equation}
In $H^{\rm CS}$ we will have the same term as for $s_i,s_j \in \{-1,+1\}$ we have the identity:
\begin{equation}
\label{eq:identity_delta_product}
    2\delta(s_i,s_j)-1\equiv s_is_j \Longleftrightarrow \delta_{s_i,s_j} \equiv \frac{s_is_j+1}{2}
\end{equation}

\subsection{Three-body terms}
Following our homogeneous mean-field hypothesis, we can write the three body terms appearing in $H^{\rm BS}$ as:
\begin{align}
\label{eq:three_body_broken}
s_is_js_k =&\left[\langle s_i \rangle + (s_i - \langle s_i\rangle)\right]\left[\langle s_j \rangle + (s_j - \langle s_j \rangle)\right]\left[\langle s_k \rangle + (s_k - \langle s_k\rangle)\right] \nonumber\\
\simeq & m^2(s_i+s_j+s_k)-2m^3
\end{align}
For the three-body terms appearing in $H^{\rm CS}$ we will make repeated use of Eq.~\eqref{eq:identity_delta_product} and of the following identity (which can be generalized at higher orders):
\begin{equation}
\label{eq:identity_multiple_delta}
    \delta(s_i,s_j,s_k)\equiv \delta(s_i,s_j)\delta(s_i,s_k)
\end{equation}
We have:
\begin{align}
\label{eq:three_body_conserved}
2\delta(s_i,s_j,s_k)-1 & = 2\left[\delta(s_i,s_j)\delta(s_i,s_k)\right]-1 \nonumber \\
& = 2\left[\frac{(s_is_j+1)(s_is_k+1)}{4}\right]-1 \nonumber \\
& \simeq \frac{2m(s_i+s_j+s_k)-3m^2-1}{2}
\end{align}
where from the second to the third line we have applied Eq.~\eqref{eq:mf_trick_1} and neglected second-order terms in the fluctuations.

\subsection{All orders}
Applying repeatedly what we have shown for the two and three-body terms appearing in $H^{\rm BS}$, we write the mean-field approximation for group interactions of arbitrary order $\ell$ appearing in $H^{BS}$ as:
\begin{equation}
\label{eq:prod-spin}
    \prod_{i =1}^{\ell+1} \ s_i =  m^{\ell} \sum _{i=1}^{\ell+1} s_i  - \left(\ell \ m^{\ell+1} \right) \ .
\end{equation}
For $H^{\rm CS}$, the generalization at all orders of Eq.~\eqref{eq:identity_delta_product} is:
\begin{equation}
    \bigotimes_{i=1}^n s_i\equiv \prod_{j=2}^n\delta(s_1, s_j)
\end{equation}
Using this, we write the many-body terms appearing in the Hamiltonian Eq.~\eqref{eq:all_orders_conserved_symmetry} as:
\begin{equation}
\label{eq:general-conserved-sim}
2\bigotimes_{i=1}^n s_i =  \frac{1}{2^{n-1}}\left[\sum_{\alpha=2}^{2 \left\lfloor \frac{n}{2} \right\rfloor} \prod_{i=1}^{\alpha} s_{i}+1\right]
\end{equation}
where $2 \left\lfloor \frac{n}{2} \right\rfloor$ is the largest even number smaller or equal to $n$.
For instance, if $n$ = 5, $2 \left\lfloor \frac{n}{2} \right\rfloor = 4$ and the latter equation will contain products of spins up to size $4$.
Inserting Eq.~\eqref{eq:prod-spin}---with the necessary re-mapping of the indices---into Eq.~\eqref{eq:general-conserved-sim} allows us to write explicitly the decoupling of the Kronecker delta in terms of the single nodes states and powers of the magnetization.

\subsection{Effective field and equation of state}
In both models and at all orders of interactions, under the homogeneous mean-field assumption we decouple all spins, we can thus write a fully decoupled Hamiltonian---neglecting all constant terms---:
\begin{equation}
\label{eq:decoupled_hamiltonian}
    H\simeq \sum_i H_i(s_i) = -h_{\rm eff.}\sum_i s_i
\end{equation}
For systems with magnetic fields and interactions up to order $\ell=2$ (pairwise and three-body interactions) the effective field takes values:
\begin{subequations}
\label{eq:effective_fields}
\begin{align}
    h_{\rm eff}^{\rm BS} & = h +\langle d_1 \rangle J_1 m + \langle d_2 \rangle J_2 m^2
    \\
    h_{\rm eff}^{\rm CS} & = h +\langle d_1 \rangle J_1 m + \langle d_2 \rangle J_2 m
\end{align}
\end{subequations}
Where we have applied a homogeneous mixing hypothesis, and $\langle d_{\ell}\rangle$ is the average generalized degree of order $\ell$ (number of interactions of order $\ell$ that each node participates in).
Using known results for the equation of state of an Ising paramagnet---the same results are obtained by finding the minima of the free energy---we have:
\begin{equation}
\label{eq:ising_paramagnet}
    m=\tanh \left[\left(\beta h_{\rm eff}(m)\right)\right]
\end{equation}
We can solve this equation numerically by fixed point iteration:
\begin{equation}
\label{eq:iterative_solution}
    m_{(t+1)} =\tanh \left[\left(\beta h_{\rm eff}(m_{(t)})\right)\right]
\end{equation}
or using other methods such as the Newton-Raphson algorithm.

\subsection{Constrained free energy}
We have shown that for both systems we can write the Hamiltonian in the mean-field approximation as:
\begin{equation}
\label{eq:hamiltonian_mf}
    H \simeq -h_{\rm eff} \sum_i s_i = -Nh_{\rm eff}m=H(m)
\end{equation}
We can then write the partition function as:
\begin{equation}
\label{eq:mf_partition_function}
    Z_n=\sum_{\mathbf{s}}\exp[-\beta H(\mathbf{s})]=\sum_m g(m)\exp[-\beta H(m)]=\sum_m \exp[-\beta H(m)+\log g(m)]
\end{equation}
where $m$ take $N+1$ possible values $-1, -1+2/N,...,1-2/N,1$ and $g(m)$ counts the number of configurations with magnetization $m$.
To compute $g(m)$ we introduce the two variables $N_+$ and $N_-$, corresponding respectively to the total number of spins taking values $+1$ and $-1$ in a given configuration of the system.
We have:
\begin{equation}
\label{eq:Nplus_Nminus}
\begin{cases}
N_++N_-=N \\
N_+-N_-=Nm
\end{cases}
\Longrightarrow
\begin{cases}
N_+=N\frac{m+1}{2} \\
N_-=N\frac{m-1}{2}
\end{cases}
\end{equation}
We can then write:
\begin{equation}
\label{eq:g_m}
    g(m)=\frac{N!}{N_+!N_-!}
\end{equation}
Using Stirling's approximation we can write:
\begin{equation}
\label{eq:stirling_log_g_m}
\log g(m) \simeq -N\left(\frac{1+m}{2}\log \frac{1+m}{2}+\frac{1-m}{2}\log \frac{1-m}{2}\right)
\end{equation}
We can then introduce the constrained free energy density:
\begin{equation}
\label{eq:constrained_free_energy_density}
    f(m)=\frac{H(m)}{N}-\frac{\log g(m)}{\beta N}
\end{equation}
and write the partition function as:
\begin{equation}
    Z_n=\sum_m \exp[-N\beta f(m)])
\end{equation}

\section{Equation of state and free energy landscape for the BS model}
\begin{figure}[H]
    \centering
    \vspace{1em}
    \includegraphics{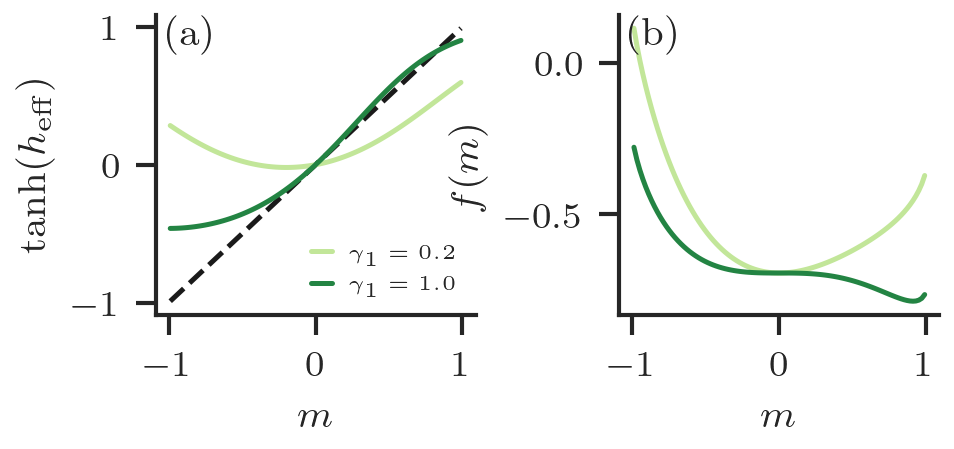}
    \caption{(a) Shape of the right-hand side of Eq.~\eqref{eq:ising_paramagnet} and (b) functional form of the mean-field constrained free energy density in the case in which $h=0$, $\ell_{\rm max}=2$, and $\gamma_2$ for the $p$-spin model.}
    \label{fig:figS1}
\end{figure}

We show in Fig.~\ref{fig:figS1} the graphical solution of the equation of state Eq.~\eqref{eq:ising_paramagnet} (panel a) and the shape of the constrained free energy density in the case $h=0$ and $\ell_{\rm max}=2$.
Depending on the values of the control parameters $\gamma_1$ and $\gamma_2$ we find the system in two states---corresponding to the stable solutions of Eq.~\eqref{eq:ising_paramagnet} and to global minima of $f(m)$---, a disordered one with $m=0$ and an ordered one with $m>0$.
Unlike the CS model, the ordered (ferromagnetic) solution is not degenerate. 
In the case in which we have even $\ell$ order of interactions---with the product of an odd number of spins appearing in the Hamiltonian---we have only one ferromagnetic solution for the equation of state, with positive magnetization.
This is because the breaking of the spin-flip symmetry introduced by the product of an odd number of terms favors the configuration where spins are aligned pointing in the $+1$ direction.

\section{Georges-Yedidia expansion}
The Georges-Yedidia (G.-Y.) expansion is a high-temperature expansion of the free-energy functional of a system performed
at fixed order parameter.
This method provides a simple and systematic way to derive corrections to mean-field theories.
In the following, we present the general setting of this methodology.
We later present the results on a general pairwise Ising model and for the CS model discussed in the main text. 
A thorough explanation of the method can be found in the original paper~\cite{georges1991expand} and in Ref.~\cite{zamponi2014meanfieldtheoryspin}. An alternative approach, focusing not on a high-temperature expansion but on a small coupling regime, was proposed in \cite{altieri2016, altieri2018} for studying constraint satisfaction problems.

Given a general spin system with Hamiltonian $H$, we construct a magnetization-dependent free energy:
\begin{equation}
    -\beta F[\mathbf{m}]=\mathcal{F}^{\beta}[\mathbf{m}]=\log \sum_{\{\mathbf{s}\}}\exp \left[-\beta H(\mathbf{s})+\sum_i \rho_i^{\beta}(s_i-m_i)\right]
\end{equation}
where $\mathbf{m}$ is a magnetization configuration of the system.
The Lagrange multipliers $\rho_i^{\beta}$ fix the magnetization at each site $i$ to their 
thermal expection $m_i=\langle s_i \rangle$. 
Such multipliers are set by the condition $\langle s_i-m_i\rangle$, we thus have:
\begin{equation}
\label{eq:expression_lambda}
\rho_i^{\beta}=-\frac{\partial}{\partial m_i}\mathcal{F}^{\beta}[\mathbf{m}]
\end{equation}
The G.-Y. expansion then consists in writing $\mathcal{F}^{\beta}[\mathbf{m}]$ using a Taylor expansion around $\beta=0$:
\begin{equation}
    \left. \mathcal{F}^{\beta}[\mathbf{m}]=\mathcal{F}^{\beta}[\mathbf{m}]\right|_{\beta=0}+\left.\frac{\partial \mathcal{F}^{\beta}[\mathbf{m}]}{\partial \beta}\right|_{\beta = 0}\beta + \frac{1}{2}\left.\frac{\partial^2 \mathcal{F}^{\beta}[\mathbf{m}]}{\partial \beta^2}\right|_{\beta = 0}\beta^2+...
\end{equation}
The derivatives of $\mathcal{F}^{\beta}[\mathbf{m}]$ evaluated at $\beta=0$ can be easily computed as spins are 
uncorrelated at infinite temperature: $\langle \prod_{i=1}^{\alpha} s_i\rangle|_{\beta=0}=\prod_{i=1}^{\alpha} m_i$.

To compute the derivatives appearing in the Taylor expansion it is convenient to define the observable:
\begin{equation}
U(\mathbf{s},\mathbf{m})=H(\mathbf{s})-\langle H \rangle- \sum_i \frac{\partial}{\partial \beta}\rho_i^{\beta}(S_i-m_i)
\end{equation}
Since $\langle s_i-m_i\rangle=0$ for all $\beta$ we have the useful relations:
\begin{subequations}
\begin{align}
	\langle U \rangle & = 0
	\\
	\frac{\partial}{\partial \beta}\langle O \rangle & = \langle \frac{\partial}{\partial \beta} O \rangle - \langle OU \rangle
\end{align}
\end{subequations}
where the second relation holds for any observable $O$.
Using the second identity we have that:
\begin{equation}
0=\frac{\partial}{\partial \beta} m_i = \frac{\partial}{\partial \beta}\langle s_i \rangle = - \langle Us_i \rangle =
- \langle U(s_i-m_i) \rangle
\end{equation}
We then have for the first derivative:
\begin{equation}
\label{eq:first_order_general}
    \frac{\partial}{\partial \beta} \mathcal{F}^{\beta}[\mathbf{m}] = \left\langle -H(\mathbf{s})+\sum_i \frac{\partial}{\partial \beta}\rho_i^{\beta}(S_i-m_i)\right\rangle = - \langle H \rangle  
\end{equation}
and for the second derivative:
\begin{equation}
\label{eq:second_order_general}
    \frac{\partial^2}{\partial \beta^2}\mathcal{F}^{\beta}[\mathbf{m}] = -\frac{\partial}{\partial \beta} \langle H \rangle = \langle HU \rangle = \langle U^2 \rangle
\end{equation}
Finally, to compute these derivatives evaluated at $\beta=0$ we need to express the derivatives of $\rho_i^\beta$ appearing in $U$.
This is done using Eq.~\eqref{eq:expression_lambda}:
\begin{equation}
    \left. \frac{\partial}{\partial \beta} \rho_i^{\beta}\right|_{\beta=0}=-\left.\frac{\partial}{\partial \beta}\frac{\partial}{\partial m_i}\mathcal{F}^{\beta}[\mathbf{m}]\right|_{\beta=0}=\frac{\partial}{\partial m_i}\langle H\rangle_{0}
\end{equation}
where $\langle...\rangle_0$ indicates the thermal average performed at $\beta=0$.

\subsection{Summary of the G.-Y. expansion for a general Ising model}
\label{subsec:gy_ising}
We display the G.-Y. expansion on a generic pairwise Ising model with uniform magnetic field.
The results for the CS model with three body interactions will turn out to be very close to those found here.
The Hamiltonian is given by:
\begin{equation}
    H(\mathbf{s})=-\frac{1}{2}\sum_{i \neq j} J_{ij} s_i s_i - h \sum_{i} s_i
\end{equation}
Considering the zeroth order term we have:
\begin{equation}
    \left.\mathcal{F}^{\beta}[\mathbf{m}]\right|_{\beta=0}=\log \sum_{\mathbf{s}}\exp\left[\sum_i \rho_i^0(s_i-m_i)\right]=-\sum_{i} \left(\frac{1+m_i}{2}\log\frac{1+m_i}{2}+\frac{1-m_i}{2}\log\frac{1-m_i}{2}\right)
\end{equation}
For the first order term we have---using Eq.~\eqref{eq:first_order_general} and the fact that at $\beta=0$ spins are uncorrelated---:
\begin{equation}
    \left. \frac{\partial}{\partial \beta} \mathcal{F}^{\beta}[\mathbf{m}]\right|_{\beta=0}=-\langle H \rangle_0 = \frac{1}{2}\sum_{i \neq j} m_i m_j + h \sum_i m_i
\end{equation}
To compute $U_0$ appearing in Eq.~\eqref{eq:second_order_general} we have that:
\begin{equation}
    \left.\frac{\partial}{\partial \beta}\rho_i^{\beta}\right|_{\beta=0}=\frac{\partial}{\partial m_i}\langle H\rangle_0 = -\sum_{j(\neq i)} J_{ij}m_j-h
\end{equation}
Plugging this into the definition of $U$ we have:
\begin{equation}
    U_0=-\frac{1}{2}\sum_{i\neq j} J_{ij}(s_i-m_i)(s_j-m_i)
\end{equation}
which gives:
\begin{equation}
    \left. \frac{\partial^2}{\partial \beta^2}\mathcal{F}^{\beta}[\mathbf{m}]\right|_{\beta=0}=\langle U_0^2 \rangle_0 = \frac{1}{2}\sum_{i \neq j} J_{ij}^2(1-m_i^2)(1-m_j)^2
\end{equation}
Collecting these results we have:
\begin{align}
    \mathcal{F}^{\beta}[\mathbf{m}]=&-\sum_{i} \left(\frac{1+m_i}{2}\log\frac{1+m_i}{2}+\frac{1-m_i}{2}\log\frac{1-m_i}{2}\right)\nonumber
    \\
    &+\frac{\beta}{2} \sum_{i \neq j} m_i m_j + h \sum_i m_i +
    \frac{\beta^2}{4}\sum_{i \neq j} J_{ij}^2(1-m_i^2)(1-m_j)^2 + O(\beta^3)
\end{align}

\subsection{CS model with three-body interactions on a fully connected 2-hypergraph}
We consider the fully connected CS model of $N$ nodes with $\ell_{\rm max}=2$ and $h=0$---\textit{i.e.} with pairwise and three-body interactions and no magnetic field.
The Hamiltonian is given by:
\begin{equation}
    H(\mathbf{s})=-\frac{J_1}{N}\sum_{i \neq j} \left[2\delta(s_i, s_j)-1\right] - \frac{J_2}{6N^2}\sum_{i \neq j \neq k}\left[2\delta(s_i, s_j, s_k) -1\right]
\end{equation}
where as usual, we have rescaled the coupling parameters to preserve the energy's extensivity.
By using Eqs.~\eqref{eq:identity_delta_product}-\eqref{eq:identity_multiple_delta} and taking into account the fact that each pair $(i,j)$ of nodes appears in $N-2$ possible groups of three nodes when neglecting constant terms the Hamiltonian reduces to:
\begin{equation}
    H(\mathbf{s}) \simeq -\left(\frac{J_1}{2N}+\frac{J_2(N-2)}{4N^2}\right)  \sum_{i\neq j} s_i s_j
\end{equation}
which is the Hamiltonian of a fully connected ferromagnetic Ising model, with a modified coupling $\Tilde{J}=J_1+J_2/2$ for $N\to \infty$.
We can then use the results obtained in \ref{subsec:gy_ising}.
Since all the spins are equivalent (the couplings are uniform) the minimum of the free energy will have a uniform magnetization $m_i=m \; \forall i$.
We see that the $\beta^2$ term in the expansion of $\mathcal{F}^{\beta}[\mathbf{m}]$ will vanish for $N\to\infty$.
We thus have---as expected for fully connected systems where the individual couplings vanish---that the mean-field results given by the first two terms of the temperature expansion are exact.

\subsection{CS model with three-body interactions on a $d$-regular 2-hypergraph}
We consider the CS model with only three-body interactions defined on a $d$-regular 2-hypergraph---\textit{i.e.} a hypergraph with only hyperedges of size 3, where each node participates in exactly $d$ hyperedges.
We have the Hamiltonian:
\begin{equation}
    H(\mathbf{s})=-J_2\sum_{(i,j,k)}\left[2\delta(s_i, s_j, s_k) -1\right] \simeq -\frac{J_2}{2}\sum_{(i,j,k)}s_i s_j + s_i s_k + s_j s_k 
\end{equation}
where the sum runs over all triplets of nodes $(i,j,k)$ that form a hyperedge and we have as usual transformed the delta into a sum of pairwise products and neglected the constant terms.

Coming to the G.-Y., the first term in the expansion gives the usual entropy of uncorrelated spins.
For the second term in the expansion we have:
\begin{equation}
    \langle H \rangle_0 = -\frac{J_2}{2}dNm^2
\end{equation}
where we have again used the results derived in \ref{subsec:gy_ising} and the fact that the number hyperedges in a $d$-regular $\ell$-hyperedge with $N$ nodes is $dN/(\ell+1)$.
In the same way, we find for the third term in the expansion:
\begin{equation}
    \langle U_0^2 \rangle_0=\frac{J_2^2}{4}dN(1-m^2)^2
\end{equation}
Putting everything together we have the expression for the free energy:
\begin{equation}
    f(m)=\beta \frac{F(m)}{N}=\left(\frac{1+m}{2}\log\frac{1+m}{2}+\frac{1-m}{2}\log\frac{1-m}{2}\right)-
 \frac{\beta J_2d}{2}m^2-\frac{\beta^2 J_2^2d}{8}(1-m^2)^2+O(\beta^3)
\end{equation}
Considering only the first two terms gives the mean-field theory, where the critical temperature is $T_c^{\rm MF}=J_2d$.
Considering also the correction we have to expand $f(m)$ in powers of $m$, then we find the temperature that equates to zero the coefficients of $m^2$:
\begin{equation}
   d( \beta J_2)^2-2d(\beta J_2)+2=0 \, \to \beta_c J_2 = 1-\sqrt{1-\frac{2}{d}} =  
\frac{1}{d}\left(1+\frac{1}{2d}+O\left(\frac{1}{d^2}\right)\right)
\end{equation}
which gives the result shown in the main text:
\begin{equation}
	T_c = J_2d\left(1-\frac{1}{2d}+O\left(\frac{1}{d^2}\right)\right)
\end{equation}
For the spontaneous magnetization $m_0$ we expand the minimization condition of $f(m)$ for small $m$:
\begin{equation}
    \beta J_2 d m -\frac{\beta^2 J_2^2d}{2}m=\tanh^{-1}m\simeq m+\frac{1}{3}m^3
\end{equation}
this equation has three solutions $m=0,\pm m_0$, where $m = 0$ must be discarded since it corresponds to a maximum of the free energy for $T<T_c$.
The spontaneous magnetization $m_0$ is given by:
\begin{equation}
    m_0=\sqrt{3\left(\beta J_2 d - \frac{\beta^2 J_2^2d}{2}-1\right)}=\sqrt{\frac{3}{T}\left(T_c-T+\frac{J_2}{2}\left(1-\frac{J_2d}{T}\right)\right)}\simeq \sqrt{-3t}\left(1+\frac{1}{2d}\right)
\end{equation}
where $t=(T-T_c)/T_c$ is the reduced temperature of the system.

\bibliography{reference_supp}